\documentclass[sigconf]{acmart}

\AtBeginDocument{%
  \providecommand\BibTeX{{%
    \normalfont B\kern-0.5em{\scshape i\kern-0.25em b}\kern-0.8em\TeX}}}

\setcopyright{acmcopyright}
\copyrightyear{2018}
\acmYear{2018}
\acmDOI{10.1145/1122445.1122456}

\acmConference[Woodstock '18]{Woodstock '18: ACM Symposium on Neural
  Gaze Detection}{June 03--05, 2018}{Woodstock, NY}
\acmBooktitle{Woodstock '18: ACM Symposium on Neural Gaze Detection,
  June 03--05, 2018, Woodstock, NY}
\acmPrice{15.00}
\acmISBN{978-1-4503-9999-9/18/06}

\begin{document}

\title{With Registered Reports Towards Large Scale Data Curation}

\author{Steffen Herbold}
\email{herbold@cs.uni-goettingen.de}
\orcid{0001-9765-2803}
\affiliation{%
  \institution{University of Goettingen, Institute of Computer Science}
  \streetaddress{Goldschmidtstr. 7}
  \city{G\"{o}ttingen}
  \country{Germany}
  \postcode{37077}
}

%\renewcommand{\shortauthors}{Herbold et al.}

% % % The abstract is a short summary of the work to be presented in the %
% article.
\begin{abstract}
The scale of manually validated data is currently limited by the effort that
small groups of researchers can invest for the curation of such data. Within
this paper, we propose the use of registered reports to scale the curation of
manually validated data. The idea is inspired by the mechanical turk and replaces
monetary payment with authorship of data set publication.
\end{abstract}





\maketitle

\newcommand{\etal}{~\textit{et al.}}

\section{Introduction}

Data from software repositories or other sources is the backbone of many
techniques that researchers are developing, e.g., automated program
repair~\cite{Gazzola2019} and defect prediction~\cite{Hosseini2017a}. We can
divide the data into three categories: raw data that has been scraped from
repositories that is published without change, e.g.,
GHTorrent~\cite{Gousios2013}; automatically processed data in which researchers
enrich raw data through automated (heuristic) algorithms, e.g.,
with software metrics or with defect labels computed with the SZZ
algorithm~\cite{Sliwerski2005}; and manually processed data, where researchers
currate manually enriched data with additional information, e.g., by
correcting issue tracking data~\cite{Herzig2013} or manually untangling
changes~\cite{Just2014,Mills2018}.

The different kinds of data have different advantages and drawbacks. While raw
data is largest in scale, working with this data requires careful and often
time-consuming downstream processing by researchers. Automatically processed
data can often directly be used for analysis by researchers, but carries
the risk of noise in case heuristics were used. Manual data is
similar to heuristic data without noise, but often smaller in scale due to the
amount of manual effort required to enrich the data.

Within this paper, we propose the research turk as a new approach for the
scaling of research projects that require manually enriching data. The concept
of the research turk is based on the mechanical turk, a method for distributing
simple task of manual labor to a workforce~\cite{Harinarayan2007}. In comparison
to the mechanical turk, the research turk also works if expert knowledge is
required and replaces monetary payment with scientific credit in form of
authorship. Publications that want to use the research turk make use of
registered reports to define the study protocol. The pre-registration defines
which data are collected, which task researchers must solve, what the
requirements for participation are, and how much effort is required to become an
author.

The remainder of this paper is structured as follows. We briefly introduce
registered reports and the mechanical turk in Section~\ref{sec:foundations}. We
present the research turk in Section~\ref{sec:research-turk} and discuss ethical
issues in Section~\ref{sec:ethics}. We present an example for how the research
turk may be used in Section~\ref{sec:example} and discuss applications other
than manual data in Section~\ref{sec:beyond}. Section~\ref{sec:conclusion} concludes the paper.
\section{Foundations}
\label{sec:foundations}

Our proposed approach builds on two concepts, which we explain in this section:
registered reports and the mechanical turk.

\subsection{Registered Reports}

The Center for Open Science states that with a registered report
``\textit{you're simply specifying to your plan in advance, before you gather
data.}''\footnote{https://cos.io/prereg/} The pre-registration of the report
defines the study protocol before the actual research project is conducted.
While this concept is not yet established in software engineering research,
there are already first results from other disciplines that demonstrate that
registered reports can improve the quality of scientific
results~\cite{Allen2018}. To the best of our knowledge, the first chance for
registered reports in software engineering will be a special track at the
International Conference for Mining Software Repositories
2020\footnote{https://2020.msrconf.org/track/msr-2020-Registered-Reports} in
cooperation with the Empirical Software
Engineering\footnote{https://link.springer.com/journal/10664} journal.

\subsection{The Mechanical Turk}

Amazon conceptualized the idea of the mechanical turk to crowd source decision
making in the internet~\cite{Harinarayan2007} with the
MTurk\footnote{https://www.mturk.com/}. The MTurk describes two different kinds
of users: \textit{Employers} and \textit{Turkers}. The \textit{Employers} post
\textit{Human Intelligence Tasks} (HITs) on the MTurk website. Such a task
could, e.g., be the identification of all images that contain a certain object.
The Employers define a payment rate for the fulfillment of the HIT. Turkers can
browse among the HITs and perform the tasks to earn money. To ensure quality of
the results, the same HIT can be solved by multiple Turkers\footnote{https://blog.mturk.com/tutorial-understanding-hits-and-assignments-d2be35102fbd}.

A key requirement for effective work with a platform like the MTurk is that
the HITs do not require expert knowledge. The ideal tasks for HITs are too
complex for state of the art algorithms, but simple for humans.
Unfortunately, this means that such a crowd working approach can only be used in
a limited way to create research data. For example, it would be highly unlikely
to find Turkers that could untangle commits~\cite{Just2014, Mills2018} or manually label bug reports~\cite{Herzig2013}.
Moreover, people that have this kind of expert knowledge would command
relatively high rates for solving the tasks, which would be economically
challenging for most research projects.

\section{The Research Turk}
\label{sec:research-turk}

The research turk is our approach to create a mechanical turk that would work
for research tasks that require expert knowledge and that can, therefore, not be
performed by the general public but only by researchers. The general concept for
the research turk is the same as for the mechanical turk, we only replace the
employer with a Principle Investigator (PI)\footnote{While we use the singular
of the term PI in the following, this may also be a group of researchers.} and
the turkers with researchers:
the PI defines HIT as research task which shall be solved by qualified
researchers. However, there are two underlying problems with the research turk:
\begin{itemize}
  \item How and where does the PI define the HIT?
  \item How are researchers motivated to solve the HIT?
\end{itemize}

In the following, we describe how registered reports can solve both
problems, enable meta analysis of the work conducted in a research turk project,
and why registered reports are better than simply advertising for the research
turk on a website.

\subsection{Definition of the HIT}

Registered reports provide the means to define the HIT, i.e., the PI would
pre-register the research and define the research protocol in advance.
The description of the research protocol for a research turk project must cover
the following aspects.
\begin{itemize}
  \item A description of the HIT that shall be solved.
  \item A strong motivation for the research project and the importance of the
  HIT for the community, including the reasons why the PI believes that a
  sufficient amount of researchers contribute by solving the HIT.
  \item An estimation of the effort for solving the HITs that demonstrates the need for
  distributing the task within the community and allows researchers to estimate
  the effort their contribution would require.
  \item A description of how the PIs ensure the quality of the results produced
  by the participating researchers.
  \item A description of when and where the data will be shared publicly and
  under which license. This description shall also contain information what
  happens with the data in case not enough researchers participate in the crowd
  working to fully address the research proposal.
\end{itemize}

Because there is no direct relationship between the PI and the researchers that
conduct the HITs, the quality assurance of the results is an important aspect
that should not be neglected. The PIs should define sanity checks for produced
data, e.g., by spot checking the solution of the HITs by the PIs or by requiring
multiple solutions for each HIT.

\subsection{Motivating Researchers}

Motivation can be provided to researchers by giving them the chance to become
authors of the study by participating in the HIT. The path to authorship needs to be
defined in the pre-registered research protocol. The venue where the report
is registered must allow for changes in the group of authors between the
registration of the report and submission of the results. To address this, the
research protocol must cover the following aspects.
\begin{itemize}
  \item A description of how researchers can sign up for the participation in the
  registered report. The registration procedure should cover potential data privacy
  and ethical concerns, e.g., which data from researchers is stored as part of
  solving the HIT, including why this data is required and how the data will be
  used.
  \item A description of the minimal level of experience required to solve the
  HITs, e.g., specified by an academic degree, years of experience, or
  similar measures. This may also include exclusion criteria that the PIs may
  apply to reject requests for participation.
  \item A description of the minimal requirements for becoming authorship of the
  report, e.g., how many HITs must be solved by a researcher.
  \item A description of what happens if researchers participate, but do
  not meet the minimal requirements for co-authorship. This could, e.g., result
  in an acknowledgement.
  \item A description of how the order of authors will be determined,
  e.g., alphabetically or based on the number of HITs that are solved.
  \item The starting date and the end date for the crowd working phase. 
\end{itemize}

We believe that offering authorship is an effective incentive for a
sufficiently large amount of researchers to contribute by solving HITs, if the
venue where the results will be published is respected and the expected impact
of the resulting publication is high. For graduate students such a way to
authorship of publications can be highly interesting, e.g., for
improving their chances of a better grade, to improve their CV, to gain
experience in conducting research tasks, or because they would directly benefit
from the outcome of the research because they can exploit the results produced.
Senior researchers may also participate, but we believe that this only happens
if they would directly benefit from the outcome of the research, e.g., because
they planned similar or related projects themselves.

\subsection{Gamification}

Prescribing minimal criteria for authorship carries the risk that researchers
only solve exactly that amount of HITs, even though they could contribute more
to the project by solving additional HITs. While this is a completely valid
approach and we cannot fault any researcher for acting that way, many research
projects would benefit from researchers solving more than the minimal
amount of HITs. We believe that there is a gamification element that may be used
for this purpose. The simplest approach would be to show a (potentially
anonymized or pseudonymized) ranking of researchers showing who currently solved the most HITs. This may motivate
researchers to raise ``higher'' in the ranking.

Such a ranking approach may directly be combined with the ordering of the author
list. Many publications order the authors by some measure of their contribution
to the work. The more someone contributed, the higher the ranking. Thus, researchers
may be motivated to solve more HITs, if they move up in the author list and,
thereby, gain a higher level of visibility.

\subsection{Meta Analysis}
\label{sec:meta-analysis}

While the solution of the HITs is the main objective of a study that
is performed using the research turk, there are also opportunities for a meta
analysis of the results. For example, the meta-analysis may cover the following
aspects.
\begin{itemize}
  \item Descriptive statistics about the number of HITs solved by each researcher.
  \item Demographic analysis, e.g., regarding the educational background of the
  researchers, their prior experience, or their geographical location.
  \item Relations of the above with the results of the quality assurance
  measures, e.g., between the number of HITs solved and the outcome of the
  quality assurance.
\end{itemize}
All meta-analysis that shall be performed should
be specified as part of the registered report.

\subsection{Advantages of Registered Reports}

The research turk could also be used without registered reports, e.g., by
creating a website that specifies the research protocol. However, we believe
that using registered reports provides multiple advantages over the website
approach.

The registered report may already be peer reviewed at the time of the
registration. This can help to ensure both the importance of the project, as
well as to avoid invalid results due to mistakes in the design of the HIT or the
evalatution procedure. Thus, researchers have an assurance that the research
protocol of the research turk study is valid and will not be rejected later due
to mistakes. This removes a big risk from the participants side that the effort
they invest does not lead to the expected payoff, i.e., the contribution to a
successful research project and the authorship of the related publication.

Moreover, the pre-registration assures that there is a public and unchangeable
record of the research protocol, which may be important in case there are
conflicts. By defining the rules for the research turk
in a peer reviewed and public forum, the PIs put their reputation at stake if
they try to change rules after starting the study or to unfairly exclude
researchers from authorship. On the other hand, the PI has the assurance that
researchers cannot unfairly claim authorship, even though they do not meet the
previously defined criteria. Thus, the registered report is similar to a
contract between the PI and the researchers. 

\section{Ethical Considerations}
\label{sec:ethics}

Since a concept like the research turk involves human participants, there are
important ethical considerations that affect the quality of the scientific work,
the use of authorship as ``payment'', and potential side-effects of such a way
to become author of scientific publications.

First, we have to consider whether offering authorship for
contributing a specified amount to a research project may lead to rushed or
invalid results, because researchers may participate with less care in such a
setting. This is a general question of any research project and, in the end,
must be handled responsibly by the PI. The PI must ensure that there are
sufficient measures for ensuring the quality of the registered report, e.g., by
letting each HIT be solved multiple times and using majority votes and/or doing
a meta study to assess threats to the validity of the results. How the quality
will be ensured, including counter measures in case of quality problems, should
be described as part of the pre-registration.

Second, it may be unethical to offer authorship because solving HITs may not be
a sufficient contribution to a research project that warrents authorship. The
IEEE defines three criteria for authorship.
``\textit{(1) Made a significant intellectual contribution to the theoretical
development, system or experimental design, prototype development, and/or the
analysis and interpretation of data associated with the work contained in the
article; (2) Contributed to drafting the article or reviewing and/or revising it for
intellectual content; (3) Approved the final version of the article as accepted for publication,
including references.}''\footnote{https://journals.ieeeauthorcenter.ieee.org/become-an-ieee-journal-author/publishing-ethics/definition-of-authorship/}
The ACM adds one additional criterion.
(4) ``\textit{Agree to be held accountable for any issues relating to correctness or integrity of the work.}''\footnote{https://www.acm.org/publications/policies/authorship}

The criterion (1) is fulfilled, because the solving of the HIT contributes to
the ``analysis and interpretation of data''. The criteria (2) and (3) can
be fulfilled, by explaining that one of the requirements for authorship is
a review of the draft of the report, including signing off on the final version of the report. This may
involve significant effort for the PI, i.e., distributing the draft to all
authors and incorporating all comments. However, this will only
improve the quality of the work due to the large amount of internal reviewing
effort. The criterion (4) can also be fulfilled by stating this responsibility as a
pre-requisite for participating the research project as part of the registered
report as well as by signing off on the final draft. In case there are
researchers who fail to review the draft of the report, they may only be
mentioned in the acknowledgements of the paper, but not as authors.

Third, we believe that the gamification element we propose does not add any
additional ethical considerations, other than those we have discussed above. In
the end, it is the responsibility of the PI to ensure that the order of authors
is fair. Solving more HITs means that more effort was spent, warrenting a higher
ranking. The PI has to ensure that this does not impact the quality of the
research using the same methods as we discussed at the beginning of this
section.
\section{Example}
\label{sec:example}

We use the untangling of bug fixing commits to demonstrate the potential of the
research turk. Just\etal~\cite{Just2014} and Mills\etal~\cite{Mills2018}
manually validated which part of a bug fixing commit were actually part of a bug
fix and reduced the data to these changes, thereby untangling the bug fixing
from other changes in the commit. This manual analysis improves the quality of
data for different research directions, e.g., program repair~\cite{Gazzola2019},
bug localization~\cite{Mills2018}, and defect prediction~\cite{Hosseini2017a}.

The drawback of this data is the scale: Version 1.5 of the
Defects4J data from Just\etal~\cite{Just2014} contains data for 438 bugs from 6
different projects, the data from Mills\etal~\cite{Mills2018} is for 620 bugs
from 13 different projects. We contacted the authors to get an estimate of the
effort that was required for the manual validation. Mills\etal~\cite{Mills2018}
estimate the effort per commit as 10-15 minutes.

Table~\ref{tbl:example} shows the
estimated effort in working days per researcher required for the manual
validation of the file changes for 10.000 commits. The table depicts two
estimates for different amounts of participating researchers: every commit is
validated by a single researcher and every commit is validated by three
researchers to improve the data quality. A group of five researchers would have
to work full-time for at least 125 days to create such a data set with three
validations. With a large group of 100 researchers, each researcher has less
than 10 days of work. Thus, this example shows how improbable it is that a
single group of researchers can invest the effort required for the manual
validation of such a large data set and how this is easily possible if a large
amount of people can be motivated to collaborate.

We believe that this example could attract a sufficient amount of researchers.
Defects4J has already been cited 362 times since the publication in
2014\footnote{According to Google Scholar on 2019-10-14}. If only one author
from every third publication that cited Defects4J would participate, there would
already be over 100 contributors to the research turk effort.

\begin{table}
\begin{tabular}{rrrc}
\textbf{Researchers} & \textbf{Validations} & \textbf{Commits} & \textbf{Effort} \\
\hline\hline
1 & 1 & 10,000 & 208.3--312.5 \\
5 & 1 & 2,000 & 41.7--62.5 \\
10 & 1 & 1,000 & 20.8--31.2 \\
50 & 1 & 200 & 4.2--6.2 \\
100 & 1 & 100 & 2.1--3.1 \\
\hline
5 & 3 & 6,000 & 125.0--187.5 \\
10 & 3 & 3,000 & 62.5--93.8 \\
50 & 3 & 600 & 12.5--18.8 \\
100 & 3 & 300 & 6.2--9.4 \\
\hline
\end{tabular}
\caption{Estimation of the effort per researcher for the untangling of 10,000 bug fixing commits. The effort is given in working days (8 hours per day).}
\label{tbl:example}
\end{table}

\section{Beyond Data Curation}
\label{sec:beyond}

While data curation is the motivating example we used throughout this paper, we
believe that the research turk may also be a great tool for replication studies
that compare algorithms.
For highly active research topics with multiple new approaches every year, it is
challenging to evaluate which approaches are currently the best within the state
of thre art, especially if the evaluation criteria differ between publications,
e.g., due to differences between the benchmark data sets or performance
indicators. As a consequence, new approaches are often only compared against the
``perceived'' state of the art, e.g., recent approaches or very popular
approaches that were often used by others for comparison. Due to this, it may
become unclear what the actual state of the art is. A large replication study is
then the only way to resolve this issue and identify differences in performances
between the different approaches, e.g., as by Herbold\etal~\cite{Herbold2017b}
for defect prediction. Currently, such replication studies require a large
amount of effort by a small group of researchers.

The research turk would offer another solution: a benchmark protocol for
approaches could be pre-registered as a research turk study, including a list of
algorithms from publications that need to be replicated. The protocol would need
to specify which data is used for the benchmark and how the implementations of
the algorithms should provide their results. Researchers could then participate
in this group effort by providing the implementations of the algorithms. This
effort could, e.g., be managed via pull requests on GitHub. This would enable
code reviews of the algorithms to ensure consistent quality. Alternatively, the
protocol could prescribe that each algorithm should be implemented several times
and the implementations are tested against each other. This would add another
dimension to the replication study, as it would be possible to evaluate by
experiments if the descriptions of the algorithms by publications are sufficient.

We believe that such replication efforts could attract a large number of
interested researchers, because such a project could help researchers get
started in a new research direction.

\section{Conclusion}
\label{sec:conclusion}

Large sets of manually validated data require a massive effort that cannot be
handled by research groups on their own. Within this paper, we propose the
research turk to scale research efforts. The research turk is based on the idea
that a collaborative research effort is defined using a registered report and
that researchers can participate by contributing to this research effort in a
defined way to gain authorship for that research.  In future work, we want to
instantiate the research turk for the validation of bug fixing commits we discussed as example. 

% \begin{acks}
% This work is partially funded by DFG Grant HE 7854/4-1.
% \end{acks}

\bibliographystyle{ACM-Reference-Format}
\bibliography{literature}

\end{document}